\documentclass{article}
\usepackage{spconf,amsmath,graphicx}
\usepackage{amsfonts,amssymb}

\usepackage{booktabs}
\usepackage{multirow}
\usepackage{tikz}
\usepackage{url}

\title{continuous speech separation with conformer}
\name{\begin{tabular}{c}Sanyuan Chen, Yu Wu,  Zhuo Chen,  Jian Wu, Jinyu Li, Takuya Yoshioka \\
		Chengyi Wang, Shujie Liu, Ming Zhou\sthanks{Emails: v-sanych, yuwu1, zhuc, wujian, jinyli, tayoshio, v-chengw, shujliu, mingzhou @microsoft.com}\end{tabular}}
\address{Microsoft Corporation}
\begin{document}
	\ninept
	\maketitle
	\begin{abstract}
		Continuous speech separation was recently proposed to deal with overlapped speech in natural conversations. While it was shown to significantly improve the speech recognition performance for multi-channel conversation transcription, its effectiveness has yet to be proven for a single-channel recording scenario. 
		%The separation model extracts a single speaker signal from a mixed speech. 
		This paper examines the use of  Conformer architecture in lieu of recurrent neural networks for the separation model. 
		%The self-attention mechanism of these architectures 
		Conformer allows the separation model to efficiently capture both local and global context information, which is helpful for speech separation. %Using a fixed speech recognition model to evaluate separation outputs on the LibriCSS dataset, 
		Experimental results using the LibriCSS dataset show that the Conformer separation model achieves state of the art results for both single-channel and multi-channel settings. Results for real meeting recordings are also presented, showing significant performance gains in both word error rate (WER) and speaker-attributed WER. 
		
		%, with a relative 23.5$\%$ word error rate (WER) reduction from bi-directional LSTM (BLSTM) in the utterance-wise evaluation  
		%and a 15.4$\%$ WER reduction in the continuous evaluation.
		
		% Furthermore, we demonstrate the effectiveness of Conformer on real meeting recordings and provide our insights about Conformer in the speech separation. 
	\end{abstract}
	\section{Introduction}
	\label{sec:intro}
	
	% speech separation history
	The advance in deep learning has drastically improved the accuracy and robustness of modern automatic speech recognition (ASR) systems in the past decade \cite{ li2018developing, movsner2019improving, 
		sun2019speaker, Li2020Comparison}, enabling various voice-based applications. However, when applied to acoustically and linguistically complicated scenarios such as conversation transcription \cite{watanabe2020chime,yoshioka2019advances}, the ASR systems still suffer from the performance limitation due to overlapped speech and quick speaker turn-taking, which break the usually assumed single active speaker condition.
	Additionally, the overlapped speech causes the so-called permutation problem \cite{hershey2016deep}, further increasing the difficulty of the  conversation transcription.
	
	Speech separation is often applied as a remedy for this problem, where the mixed speech is  processed by a specially trained separation network before ASR . 
	Starting from deep clustering (DC) \cite{hershey2016deep} and permutation invariant training (PIT) \cite{yu2017permutation, kolbaek2017multitalker}, various separation models have been shown effective in handling overlapped speech \cite{yoshioka2019advances,luo2019conv,luo2020dual,chen2020dual}. 
	%The network architecture has been rigorously explored for better separation capability \cite{luo2020dual,chen2020dual}. 
	Among the network architectures proposed thus far, the Transformer \cite{chen2020dual} based approach achieved a promising result. Transformer was first introduced  for machine translation \cite{vaswani2017attention} and later extended to speech processing  \cite{dong2018speech}. A Transformer based speech separation architecture was proposed in \cite{chen2020dual}, 
	achieving the state of the art separation quality on the WSJ0-2mix dataset. It was also reported in \cite{chang2020end} that incorporating Transformer into an end-to-end multi-speaker recognition network yielded higher recognition accuracy. 
	%the authors incorporate the transformer with the end to end multi-speaker recognition network, and reports a better recognition accuracy. 
	However, both studies were evaluated on artificially simulated data sets that only considered overlapped speech, assuming the utterance boundaries to be provided, which significantly differs from the real conversational transcription scenario \cite{yoshioka2019advances,chen2020continuous}.
	% In \cite{luo2019conv,luo2020dual}, the author proposed the time domain separation methods that lead to major improvements in separation audio quality. In \cite{delcroix2018single,kinoshita2018listening,wang2019speech}, the additional speaker identification module is introduced to enhance the separation performance for both perceptual and recognition. And in \cite{wang2018multi,yoshioka2018multi}, the multi-channel extension separation has been shown to be effective.
	% % On commonly evaluation dataset WSJ0-2mix\cite{hershey2016deep}, the state of the art performance in separation quality has raised from 6.3dB\cite{hershey2016deep} in signal to distortion ratio\cite{fevotte2005bss_eval} to 20.4dB \cite{chen2020dual}.
	% In \cite{yoshioka2019advances,yoshioka2019low}, the authors propose to apply the speech separation in a continuous processing manner and integrate it in a conversation transcription system that results in significant word error rate (WER) reduction on the task of real recorded meetings. 
	
	% In this work, we explore the further performance improvement on continuous speech separation, with the application in conversation transcription, by integrating the transformer and conformer. 

	In this work, 
	inspired by the recent advances in transducer-based end-to-end ASR modeling, which has evolved from a recurrent neural network (RNN) transducer \cite{he2019streaming} to Transformer \cite{zhang2020transformer} and Conformer \cite{gulati2020conformer} transducers, we examine the use of the  Conformer architecture for continuous speech separation (CSS) \cite{yoshioka2018multi}. Unlike the prior speech separation studies, in CSS, the separation network continuously receives a mixed speech signal, performs separation, and routes each separated utterance to one of its output channels in a way that each output channel contains overlap-free signals. 
	This allows a standard ASR system trained with single speaker utterances to be directly applied to each output channel to generate transcriptions.
	The proposed system is evaluated by using the LibriCSS dataset \cite{chen2020continuous}, which consists of real recordings of long-form multi-talker sessions that were created by concatenating and mixing LibriSpeech utterances with various overlap ratios.
	Our proposed network significantly outperforms the RNN-based baseline systems, achieving the new state of the art performance on this dataset. 
	%The conformer based separation system reduces the WER by a relatively 23.5$\%$ and 15.4$\%$  for the utterance-wise evaluation and continuous evaluation respectively. 
	Evaluation results on real meetings are also presented along with tricks for further performance improvement.  %If we use an end-to-end Transformer ASR model to evaluate separation systems on LibriCSS 40$\%$ overlap recordings,   %We will release our model and test code at \url{xxxxx}.
	%$in the 40$\%$ speech overlap scenario, which significantly outperforms BLSTM model result 8.9$\%$ and 12.3$\%$. We will release our model and test code at \url{xxxxx}.

	% Conformer was proposed in \cite{gulati2020conformer}, and reports the best performance in speech recognition benchmark test on Libri-speech dataset\cite{panayotov2015librispeech}. 

	% we explore the application of transformer and conformer architecture\cite{gulati2020conformer} in speech separation. 

	% Transformer based network has also been used in speech separation task in previous arts. In \cite{chen2020dual}, a transformer based speech separation architecture is proposed and achieves the state of the art separation quality on WSJ0-2mix dataset. 
	% In \cite{chang2020end}, the author incorporate the transformer with the end to end multi-speaker recognition network, and reports a better recognition accuracy. 
	
	% However, both works were evaluated on artificially simulated data set that only considers the overlapped speech and assumes the utterance boundary is provided, which significantly differs from the real world conversation recordings\cite{chen2020continuous}.

	% overview on performance of different data set
	% LibriCSS, result, problems
	% transformer and conformer
	% in this paper
	
	\section{Approach}
	\subsection{Problem Formulation}
	The goal of speech separation is to  estimate individual speaker signals from their mixture, where the source  signals may be overlapped with each other wholly or partially. The mixed signal is formulated as 
	%\begin{equation}
	%    y(t)=\sum_{s=1}^S x_s(t),
	%\end{equation} 
	$y(t)=\sum_{s=1}^S x_s(t)$, 
	where $t$ is the time index, $x_s(t)$ denotes the $s$-th source signal, and $y(t)$ is the mixed signal. 
	%Their  short-time Fourier transforms (STFT) are represented as $\mathbf{Y}(t,f)$ and $\mathbf{X}_s(t,f)$, respectively.
	Following \cite{yoshioka2018multi}, 
	when $C$ microphones are available, the model input to the separation model can be obtained as 
	\begin{equation}
	\mathbf{Y}(t,f) = \mathbf{Y}^1(t,f) \oplus \text{IPD}(2) \ldots \oplus \text{IPD}(C), \
	\end{equation} where $\oplus$ means a concatenation operation, $\mathbf{Y}^i(t,f)$ refers to the STFT of the $i$-th channel, $\text{IPD}(i)$ is the inter-channel phase difference between the $i$-th channel and the first channel, i.e. $ \text{IPD}(i) = \theta^{i}(t,f) - \theta^1(t,f)$ with $\theta^{i}(t,f)$ being the phase of $\mathbf{Y}^i(t,f)$. 
	These features are normalized along the time axis. If $C=1$, it reduces to a single channel speech separation task.

	Following \cite{wang2014training, erdogan2017deep}, a group of masks $\{\mathbf{M}_s(t,f)\}_{1 \leq s \leq S}$ are estimated with a deep learning model $f(\cdot)$ instead of $f$ directly predicting the source STFTs. 
	% The masks are constrained by $\mathbf{M}_s(t,f) \geq 0$ and $\sum_{s=1}^S \mathbf{M}_s(t,f) = 1$.
	Each source STFT, $\mathbf{X}_s(t,f)$, is obtained as $\mathbf{M}_s(t,f) \odot  \mathbf{Y}^1(t,f) $, where $\odot$ is an elementwise product. 
	For the multi-channel setting, 
	the source signals are obtained with adaptive minimum variance distortionless response (MVDR) beamforming~\cite{6516079}. 
	In this paper, we employ the Conformer  structure \cite{gulati2020conformer} as $f(\cdot)$ to estimate the masks for (continuous) speech separation. 
	
	\subsection{Model structure}
	
	\begin{figure}[t]		
		\begin{center}
			\includegraphics[width=.8\columnwidth,  trim={0cm 2cm 0cm 0cm}]{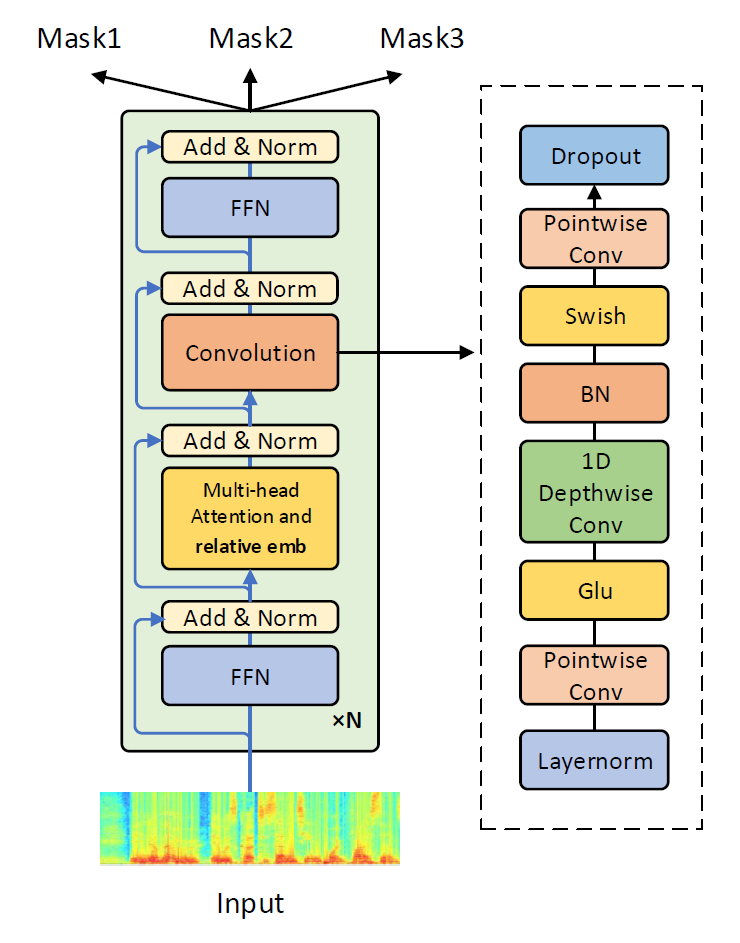}
		\end{center}
		
		\caption{Conformer architecture. There are three mask outputs, two for speakers and one for noise.}\label{fig:conformer} 
		\vspace{-4.5mm}
		
	\end{figure} 
	
	Conformer \cite{gulati2020conformer} is a state-of-the-art ASR encoder architecture, which inserts a convolution layer into a Transformer block  to increase the local information modeling capability of the traditional Transformer model \cite{vaswani2017attention}. The architecture of the Conformer is shown in Fig. \ref{fig:conformer}, where each block consists of a self-attention module, a convolution module, and a macron-feedforward module. A chunk of $ \mathbf{Y}(t,f)$ over time frames and frequency bins is the input of the first block. Suppose that the input to the $i$-th block is $z$, the $i$-th block output is calculaed as
	\begin{flalign}
	\hat{z} & = z + \frac{1}{2}\text{FFN}(z) \\
	z' & = \text{selfattention} (\hat{z}) + \hat{z}  \\
	z'' & = \text{conv} (z') + z'\\
	output &= \text{layernorm} (z'' + \frac{1}{2}\text{FFN}(z'')),
	%\mathcal{L}_{FMLM} & = - \sum_{\tilde{y}^s_i \in \tilde{\bm{y}}^s} \text{log} p(\tilde{y}^s_i|\tilde{\bm{x}})
	\end{flalign} where
	$\text{FFN}()$, $\text{selfattention}()$, $\text{conv}()$, and $\text{layernorm}()$ denote the feed forward network, self-attention module, convolution module, and layer normalization, respectively. 
	In the self-attention module, $\mathbf{\hat{z}}$ is linearly converted to $\mathbf{Q,K,V}$ with three different parameter matrices. Then, we apply a multi-head self-attention mechanism
	
	\begin{flalign} 
	\text{Multihead}(\mathbf{Q,K,V}) &= [\mathbf{H_1} \ldots \mathbf{H}_{d_{head}}]\mathbf{W}^{head}\\
	%\text{where~}  
	\mathbf{H_i} &=\text{softmax}(\frac{\mathbf{Q_i}(\mathbf{K_i+pos})^\intercal}{\sqrt{d_k}})\mathbf{V_i\label{self_att}}, %\\ 
	%  \nonumber
	\end{flalign} where  $d_k$ is the dimensionality of the feature vector, $d_{head}$ is the number of the attention heads. $\mathbf{pos} = \{rel_{m,n}\} \in \mathbb{R}^{M \times M \times d_k}$ is the relative position embedding \cite{shaw2018self}, where $M$ is the maximum chunk length and $rel_{m,n} \in \mathbb{R}^{d_k}$ is a vector representing the offset of $m$ and $n$ with $m$ and $n$ denoting the $m$-th vector of $\mathbf{Q_i}$ and the $n$-th vector of $\mathbf{K_i}$, respectively.  
	The Convolution  starts with  a pointwise convolution and a gated linear unit (GLU), followed by a 1-D depthwise convolution layer with a Batchnorm \cite{ioffe2015batch} and a Swish activation. After obtaining the Conformer output, we further convert it to a mask matrix as $  \mathbf{M}_s(t,f) = \text{sigmoid}( \text{FFN}_s(output))$.

	% The relative embedding $rel_{m,n} \in \mathbb{R}^d$ considers offset of $q_m$ and $k_n$ in attention weight computation. $\forall m,n$, the un-normalized weight of $m$-th key turns from $q_m \cdot k_n^\intercal$ to $q_m \cdot k_n^\intercal + q_m \cdot rel_{m-n}^\intercal$

	\subsection{Chunk-wise processing for continuous separation}
	\label{ssec:sliding-window}
	The speech overlap usually takes place in a natural conversation which may last for tens of minutes or longer. To deal with such long input signals, CSS generates a predefined number of signals where overlapped utterances are separated and then routed to different output channels.

\begin{figure}[t]
		\centering
		\begin{tikzpicture}
		\draw (0,0 ) node[inner sep=0] {\includegraphics[height=2.5cm, clip]{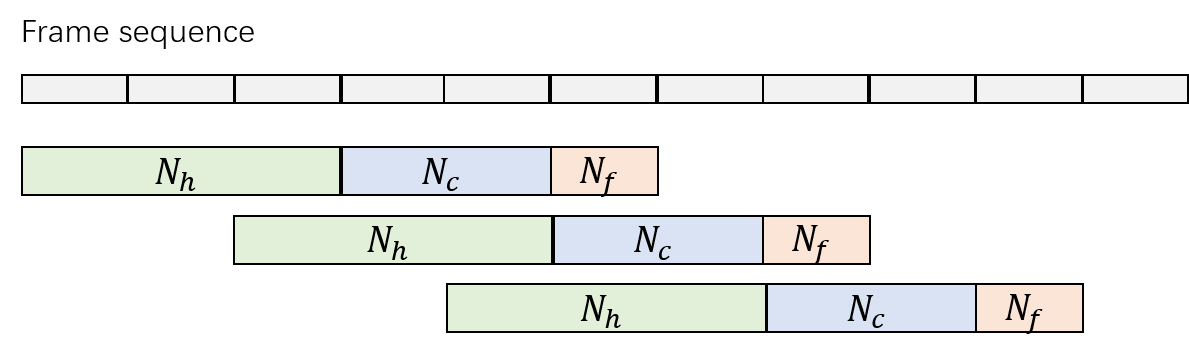}};
		\end{tikzpicture}
		\caption{Chunk-wise processing is employed to enable streaming processing for continuous speech separation.
		}\label{fig:sliding_window}
		\vspace{-4.5mm}
	\end{figure}

	To enable this, we employ the chunk-wise processing proposed in \cite{Yoshioka2018Unmix} at test time. 
	A sliding-window is applied as illustrated in Figure~\ref{fig:sliding_window}, which contains three sub-windows, representing the history ($N_h$ frames), the current segment ($N_c$ frames), and the future context ($N_f$ frames).  
	%The models are trained on the split chunks $N=[N_h \oplus N_c \oplus N_f]$. 
	We move the window position forward by $N_c$ frames each time, and compute the masks for the current $N_c$ frames using the whole $N$-frame-long chunk. 
	% Regarding the overlap region of adjacent windows, we average their mask matrices for the beamforming process followed. 
	
	To further consider the history information beyond the current chunk, we also consider taking account of the previous chunks in the self-attention module. Following Transformer-XL \cite{dai2019transformer}, the Equation \ref{self_att} is rewritten as
	\begin{equation}
	\text{softmax}(\frac{\mathbf{Q_i}(\mathbf{K_i \oplus K_{cache,i}+pos})^\intercal}{\sqrt{d_k}})(\mathbf{V_i \oplus V_{cache,i}})
	\end{equation}
	where $\mathbf{Q}$ is obtained by the current chunk while $\mathbf{K}$ and $\mathbf{V}$ are the concatenations of the previous and current changes in the key and value spaces, respectively. The dimensionality of $\mathbf{K_{cache,i}}$ depends on the number of the history chunks considered. 
	
	%The separated signals are generated either by  spectral masking or mask-based adaptive minimum
	%variance distortionless response (MVDR) beamforming \cite{yoshioka2019advances}. 
	%Regarding the overlap region of adjacent windows, we average their mask matrices for the beamforming.
	%PIT \cite{yu2017permutation} is employed to minimize the Euclidean distance between the reference and predicted signals. 
	
	\section{EXPERIMENT}
	
	\begin{table*}[!t]
		\centering
		\caption{ Utterance-wise evaluation for seven-channel and single-channel settings. Two numbers in a cell denote \%WER of the \textbf{hybrid ASR model} used in LibriCSS \cite{chen2020continuous} and \textbf{E2E Transformer} based ASR model \cite{wang2019semantic}. 0S and 0L are utterances with short/long inter-utterance silence.}
		\label{tab:7ch_utt_result}
		
		\begin{tabular}{l|cccccc} 
			\toprule \hline
			\multirow{2}{*}{\textbf{System}} &
			\multicolumn{6}{c}{\textbf{Overlap ratio in \%}} \\ &
			0S & 0L & 10 & 20 & 30 & 40   \\ \hline
			No separation \cite{chen2020continuous} & 11.8/5.5 & 11.7/5.2 & 18.8/11.4 & 27.2/18.8 & 35.6/27.7 & 43.3/36.6 \\\hline
			&   \multicolumn{6}{c}{Seven-channel Evaluation} \\ 
			\hline
			% 		BLSTM \cite{chen2020continuous} & 8.4/3.5 & 8.3/3.6 & 11.6/4.5 & 15.8/5.9 & 18.7/8.4 & 21.7/10.0 \\
			BLSTM & \textbf{7.0}/\textbf{3.1} & 7.5/\textbf{3.3} & 10.8/4.3 & 13.4/5.6 & 16.5/7.5 & 18.8/8.9 \\ 
			% 		Transformer-base & 8.5/3.5 & 8.9/3.6 &	11.4/4.0 &	13.1/\textbf{4.8} &	14.6/6.1 &	16.4/6.6 \\
			Transformer-base & 8.3/3.4 & 8.4/3.4 & 11.4/4.1 & 12.5/4.8 & 14.7/6.4 & 16.9/7.2 \\					
			Transformer-large & 7.5/\textbf{3.1} &	7.7/3.4 &	10.1/\textbf{3.7} &	12.3/\textbf{4.8} &	14.1/5.9 &	16.0/6.3 \\
			
			Conformer-base & 7.3/\textbf{3.1} & \textbf{7.3}/\textbf{3.3} & \textbf{9.6}/3.9 & 11.9/\textbf{4.8} &	13.9/6.0 &	15.9/6.8 \\
			Conformer-large & 7.2/\textbf{3.1} & 7.5/\textbf{3.3} & \textbf{9.6}/\textbf{3.7} & \textbf{11.3}/\textbf{4.8} & \textbf{13.7}/\textbf{5.6} & \textbf{15.1}/\textbf{6.2} \\ \hline
			&   \multicolumn{6}{c}{Single-channel Evaluation}  \\ \hline
			
			BLSTM &  15.8/6.4 & 14.2/5.8 & 18.9/9.6 & 25.4/15.3 & 31.6/20.5 & 35.5/25.2 \\
			
			% 		Transformer-base & 13.7/5.6 & 12.9/5.5 & 17.1/8.5 & 22.4/12.6 & 27.3/16.7 & 32.2/20.4 \\
			Transformer-base & 13.2/5.5 & 12.3/5.2 & 16.5/8.3 & 21.8/12.1 & 26.2/15.6 & 30.6/19.3 \\
			Transformer-large & 13.0/\textbf{5.3} & 12.4/5.1 & 15.5/\textbf{7.4} & 20.1/11.1 & 24.6/\textbf{13.5} & 27.9/\textbf{17.0} \\
			
			Conformer-base & 13.8/5.6 & 12.5/5.4 & 16.7/8.2 & 21.6/11.8 & 26.1/15.5 & 30.1/18.9 \\
			Conformer-large & \textbf{12.9}/5.4 & 12.2/\textbf{5.0} & \textbf{15.1}/7.5 & \textbf{20.1}/\textbf{10.7} & \textbf{24.3}/13.8 & \textbf{27.6}/17.1 \\ \hline
			\bottomrule
		\end{tabular}
		% \vspace{-1mm}
	\end{table*}
	\subsection{Datasets}
	Our training dataset consists of 219 hours of artificially reverberated and mixed utterances that sampled randomly from WSJ1 \cite{wsj1}. Four different mixture types described in \cite{yoshioka2018multi} are included in the training set. To generate each training mixture, we randomly pick one or two speakers from  WSJ1 
	and convolve each with a 7 channel room impulse response (RIR) simulated with the image method \cite{imagemethod}. The reverberated signals are then rescaled and mixed with a source energy ratio  between -5 and 5 dB. In addition, we add simulated isotropic noise \cite{habets2007generating} with a 0--10 dB signal to noise ratio. The average overlap ratio of the training set is around 50\%. 
	
	LibriCSS is used for evaluation \cite{chen2020continuous}. 
	The dataset has 10 hours of seven-channel recordings of mixed and concatenated LibriSpeech test utterances. The recordings were made by playing back the mixed audio in a meeting room. 
	Two evaluation schemes are used: utterance-wise evaluation and continuous input evaluation. 
	In the former evaluation, the long-form recordings are segmented into individual utterances by using ground-truth time marks to evaluate the pure separation performance. 
	In the contuous input evaluation, systems have to deal with the unsegmented recordings and thus CSS is needed. 
	
	%It should be noted that we don't apply the original training set from LibriCSS\cite{chen2020continuous}, as we would like to see the potential impact from the data set mismatch. We observe that the model trained on simulation from WSJ1 resulted in similar performance as ones using Librispeech as source data. 
	
	% 206 hours of artificially reverberated and mixed speech signals as the training data, and 
	
	%We test our model performance under a one-channel setting and a seven-channel setting. We conducted both the utterance-wise evaluation and continuous input evaluation. 

	\subsection{Implementation details}
	We use BLSTM and Transformers as our baseline speech separation models.
	The BLSTM model has three BLSTM layers with 1024 input dimensions and 512 hidden dimensions, resulting in 21.80M parameters. There are three masks, two for speakers and one for noise. The noise mask is used to enhance the beamforming \cite{Yoshioka2018Unmix}.
	We use three sigmoid projection layers to estimate each mask.  Transformer-base and Transformer-large models with 21.90M and 58.33M parameters are our two Transformer-based baselines. The Transformer-base model consists of 16 Transformer encoder layers with 4 attention heads, 256 attention dimensions and 2048 FFN dimensions. The Transformer-large model consists of 18 Transformer encoder layers with 8 attention heads, 512 attention dimensions and 2048 FFN dimensions.
	
	As with the Transformer baseline models, we experiment with two Conformer-based models, 
	Conformer-base and Conformer-large. They have 22.07M and 58.72M parameters, respectively. The Conformer-base model consists of 16 Conformer encoder layers with 4 attention heads, 256 attention dimensions and 1024 FFN dimensions. The Conformer-large model consists of 18 Conformer encoder layers with 8 attention heads, 512 attention dimensions and 1024 FFN dimensions. 
	Both Conformer and Transformer are trained with the AdamW optimizer \cite{loshchilov2018decoupled}, where the weight decay is set to 1e-2.
	We set the learning rate to 1e-4 and use a warm-up learning schedule with a linear decay, in which the warmp-up step is 10,000 and the training step is 260,000. 
	
	We use two ASR models to evaluate the speech separation accuracy. One is the ASR model used in the original LibriCSS publication \cite{chen2020continuous}, which is a hybrid system using a BLSTM acoustic model and a 4-gram language model. The other one is one of the best open source end-to-end Transformer  based ASR models \cite{wang2019semantic}, which achieves 2.08\% and 4.95\% word error rates (WERs) for LibriSpeech test-clean and test-other, respectively. Following \cite{chen2020continuous}, we generate the separated speech signals with  spectral masking and mask-based adaptive minimum variance distortionless response (MVDR) beamforming for the single-channel and seven-channel cases, respectively. 
	For a fair comparison, we follow the LibriCSS setting for chunk-wise CSS processing, where $N_h$, $N_c$, $N_f$ are set to 1.2s, 0.8s, 0.4s respectively.
	
	% \subsection{Results}
	
	\subsection{Results for  utterance wise evaluation}
	
	Table~\ref{tab:7ch_utt_result} shows the WER of the utterance wise evaluation for the seven-channel and single-channel settings. 
	Our Conformer models achieved state-of-the-art results. Compared with BLSTM,  Conformer-base
	yielded substantial WER gains for the 7-channel setting.
	%obtained about a 15.4$\%$ relative gain for the large-overlap setting (the one with an overlap ratio of 40$\%$) with the hybrid ASR system. 
	%With E2E Transformer model, it  achieved a 23.5$\%$ WER gain, demonstrating the superiority of the self-attention mechanism. 
	The fact that 
	the Conformer-base model outperformed Transformer-base for almost all the settings indicates  Conformer's superior local modeling capability. Also, the larger models achieved better performance in the highly overlapped settings. 
	%Conformer-large was better than Conformer-base by 8$\%$ in the 40$\%$-overlap separation task. 
	As regards the single-channel case, 
	while the overall WERs were higher, 
	% The result degraded significantly if only one channel is used, indicating the importance of multi-channel microphone. 
	the trend was consistent between the single- and multi-channel cases, except for the non-overlap scenario. With the seven channel input, all models showed similar performance for 0S and 0L.
	On the other hand, when only one channel was used,  
	the self-attention models were markedly better. This could indicate that the seven-channel features contain sufficiently rich information for simpler networks to do the beamforming well. Meanwhile, the information in the single-channel signal is quite limited, requiring a more advanced structure.

	\subsection{Results for continuous input evaluation}
	
	\begin{table*}[!t]
		\centering
		
		\caption{ Continuous speech separation evaluation for seven-channel and single-channel settings. }
		\label{tab:7ch_cont_result}
		
		\begin{tabular}{l|cccccc}
			\toprule \hline
			\multirow{2}{*}{\textbf{System}} &
			\multicolumn{6}{c}{\textbf{Overlap ratio in \%}} \\ &
			0S & 0L & 10 & 20 & 30 & 40   \\ 
			\hline
			No separation \cite{chen2020continuous} & 15.4/12.7 & 11.5/5.7 & 21.7/17.6 & 27.0/24.4 & 34.3/30.9 & 40.5/37.5 \\\hline
			&   \multicolumn{6}{c}{Seven-channel Evaluation} \\ 
			\hline
			% 		BLSTM \cite{chen2020continuous} & 8.4/3.5 & 8.3/3.6 & 11.6/4.5 & 15.8/5.9 & 18.7/8.4 & 21.7/10.0 \\
			BLSTM & 11.4/6.0 & \textbf{8.4}/4.1 & 13.1/7.0 & 14.9/7.9 & 18.7/11.5 & 20.5/12.3 \\ 
			% 		Transformer-base & 8.5/3.5 & 8.9/3.6 &	11.4/4.0 &	13.1/\textbf{4.8} &	14.6/6.1 &	16.4/6.6 \\
			Transformer-base & 12.0/5.6 & 9.1/4.4 & 13.4/6.2 & 14.4/6.8 & 18.5/9.7 & 19.9/10.3 \\					
			Transformer-large & \textbf{10.9}/5.4 & 8.8/\textbf{4.0} & \textbf{12.6}/6.0 & 13.6/\textbf{6.7} & \textbf{17.2}/9.3 & \textbf{18.9}/10.2 \\
			
			Conformer-base & 11.1/5.6 & 8.7/\textbf{4.0} & 12.8/6.1 & 13.8/\textbf{6.7} & 17.6/9.4 & 19.6/10.4 \\
			Conformer-large & 11.0/\textbf{5.2} & 8.7/\textbf{4.0} & \textbf{12.6}/\textbf{5.8} & \textbf{13.5}/6.8 & 17.6/\textbf{9.0} & 19.6/\textbf{10.0} \\ 
			% Conformer$_{xl}$-base &  11.0/5.5 & \textbf{8.4}/4.2 & 12.8/6.6 & 14.2/7.5 & 17.4/9.6 & 20.1/11.0\\
			% Conformer$_{xl}$-base & 11.2/5.6 & 8.6/4.1 & 12.8/6.6 & 13.8/7.4 & 16.9/9.7 & 19.2/10.9 \\
			Conformer$_{xl}$-base & 11.4/5.4 & 8.7/4.1 & 13.2/6.2 & 13.6/6.7 & 17.8/9.5 & 20.0/10.8 \\
			Conformer$_{xl}$-large &  11.0/5.2 & 8.8/4.1 & 12.9/5.8 & 13.7/6.7 & 17.5/9.4 & 19.8/10.6\\ \hline
			&   \multicolumn{6}{c}{Single-channel Evaluation}  \\ \hline
			
			BLSTM &  19.1/11.7 & 16.1/9.7 & 22.1/14.5 & 27.4/19.1 & 33.0/25.9 & 37.6/30.1 \\
			
			% 		Transformer-base & 13.7/5.6 & 12.9/5.5 & 17.1/8.5 & 22.4/12.6 & 27.3/16.7 & 32.2/20.4 \\
			Transformer-base & 13.8/7.1 & \textbf{11.5}/6.6 & 16.7/9.6 & 20.8/13.3 & 26.7/18.6 & 31.0/21.6 \\
			Transformer-large & \textbf{13.0}/7.2 & 12.3/6.9 & \textbf{15.8}/9.5 & \textbf{19.8}/\textbf{12.2} & \textbf{25.3}/16.9 & \textbf{28.6}/\textbf{19.3} \\
			
			Conformer-base & 14.1/7.7 & 13.0/7.1 & 17.4/10.6 & 21.9/13.7 & 27.4/18.7 & 32.0/22.4 \\
			Conformer-large & 13.3/\textbf{6.9} & 11.7/\textbf{6.1} & 16.3/\textbf{9.1} & 20.7/12.5 & 25.6/\textbf{16.7} & 29.3/\textbf{19.3} \\ \hline
			
			\bottomrule
		\end{tabular}
		% \vspace{-1mm}
	\end{table*}
	
	% \begin{table}[]
	%     \centering
	%     \footnotesize
	%     \setlength{\tabcolsep}{.7mm}
	%     \begin{tabular}{lcccccc}
	%          \toprule
	% 		\multirow{3}{*}{\textbf{System}} &
	% 		\multicolumn{6}{c}{\textbf{Overlap ratio in \%}} \\
	%  &
	% 		0S & 0L & 10 & 20 & 30 & 40   \\ 
	% 		\midrule
	% 		No separation \cite{chen2020continuous} &  15.4/12.7 & 11.5/5.7 & 21.7/17.6 & 27.0/24.4 & 34.3/30.9 & 40.5/37.5 \\
	% 		\midrule
	% % 		BLSTM \cite{chen2020continuous} &  11.9/ & 9.7/ & 13.6/ & 15.0/ & 19.9/ & 21.9/ \\
	% 		BLSTM & 11.4/6.0 & \textbf{8.4}/4.1 & 13.1/7.0 & 14.9/7.9 & 18.7/11.5 & 20.5/12.3 \\
	% 		\midrule
	% % 		Transformer-base & 11.6/5.4 & 8.9/4.2 & 13.0/6.1 & 14.3/7.0 & 17.5/9.7 & 19.8/10.7 \\
	% 		Transformer-base & 12.0/5.6 & 9.1/4.4 & 13.4/6.2 & 14.4/6.8 & 18.5/9.7 & 19.9/10.3 \\
	%         Transformer-large & \textbf{10.9}/5.4 & 8.8/\textbf{4.0} & \textbf{12.6}/6.0 & 13.6/\textbf{6.7} & \textbf{17.2}/9.3 & \textbf{18.9}/10.2 \\
	%         \midrule
	%         Conformer-base & 11.1/5.6 & 8.7/\textbf{4.0} & 12.8/6.1 & 13.8/\textbf{6.7} & 17.6/9.4 & 19.6/10.4 \\
	%         Conformer-large & 11.0/\textbf{5.2} & 8.7/\textbf{4.0} & \textbf{12.6}/\textbf{5.8} & \textbf{13.5}/6.8 & 17.6/\textbf{9.0} & 19.6/\textbf{10.0} \\
	% 		\bottomrule
	%     \end{tabular}
	%     \caption{Seven-channel continues speech separation evaluation.}
	%     \label{tab:7ch_cont_result}
	% \end{table}
	
	Table~\ref{tab:7ch_cont_result} shows the continuous input evaluation results. 
	%In contrasted to the utterance-wise evaluation, all models become worse in the continuous setting, indicating the scenario is much more difficult for speech separation. 
	The Conformer and Transformer models performed consistently better than BLSTM, but their performance gap became smaller in the large overlap test-set. The relative WER gains obtained with 
	Conformer-base over BLSTM were 4$\%$ and 15$\%$ for the hybrid and transducer ASR sytems, respectively, which were smaller than those obtained for the utterance-wise evaluation. 
	A possible explanation is that the self-attention based methods are good at using global information while the chunk-wise processing limits teh use of the context information. 
	
	It is noteworthy that 0S results were much worse than those of 0L only in the continuous evaluation, which is consistent with the previous report \cite{chen2020continuous}.
	The 0S dataset contains much more quick speaker turn changes, imposing a challenge for both speech separation and ASR.  
	The self-attention-based models showed clear improvement over BLSTM, indicating that they are also helpful for dealing with turn-takings in natural conversations. 
	
	%An explanation is that a longer silence makes the task easier, so 0L enjoy better performance. It shows the quick turn poses a challenge for speech separation system. We can observe a clear improvement of self-attention models on the 0S, indicating self-attention model is not only effective on speech with overlap, but also effective on quick turn conversations. 
	
	Table~\ref{tab:7ch_cont_result} also shows that 
	the Conformer$_{xl}$ models using 
	longer context information did not result in lower WERs especially in the large overlap ratio settings. 
	Two factors may have contributed to the performance degradation. 
	1) The unexpected noise may have been introduced from the use of the longer history, which may contain more speakers' voices. 2) Also, we did not consider the overlap regions of the adjacent windows during training, possibly making the training/testing gap greater and resulting in sub-optimal performance. We leave the training with overlap regions for the future work.

	\subsection{Results on large scale real meetings}
	To further verify the effectiveness of our method, we further conduct an experiment on an internal real conversation corpus which consists of 15.8 hours of single channel recordings of daily group discussions, noted as the Real Conversation dataset. In this dataset, the per-meeting speaker number ranges from 3 to 22. We applied a modified version of the conversation transcription system of \cite{yoshioka2019advances}, where a large scale trained speech recognizer and speaker embedding extractor were included, 
	%an agglomerate hierarchical based diarization system 
	to obtain  speaker attributed transcriptions.
	%and the speaker attributed word error rate was used for evaluation. 
	
	Compared with LibriCSS, those real meetings are significantly more complex with respect to the acoustics, linguistics, and inter-speaker dynamics. To deal with the real data challenges, 
	three improvements were made. Firstly, we increased the training data amount to 1500 hours.
	Additional clean speech samples were taken from a Microsoft internal corpus and they were mixed with the simulation setup as Section 3.1. 
	Secondly, the separation network sometimes generated a low volume residual signal from the redundant output channel for single speaker regions, which increased the word insertion errors. To mitigate this, we introduced a merging scheme, where the two channel outputs were merged when a single active speaker was judged to be present. 
	%A simple energy-based merger was applied here, as suggested in eqn. \ref{eqn:merger}, 
	The merger was triggered when only one masked channel had a significantly large energy. Lastly, to reduce the distortion introduced by the masking operation, we used single speaker signals corruped by background noise as a training target. This allowed the separation network to focus only on the separation task and leave the noise to the ASR model. 
	The WER and speaker attributed WER (SA-WER) were used for evaluation, where the latter assesses the combined quality of speech transcription and  speaker diarization \cite{yoshioka2019advances}.

	\begin{table}[!h]
		\centering
		\caption{Continuous evaluation on a real meeting dataset.}
		\label{tab:prod}
		% \footnotesize
		
		%     (array([ 0.        ,  0.06461658,  0.07220077,  0.07184241, -0.10788382,
		%         -0.11767442, -0.15640038]),
		%  array([ 0.        ,  0.18809832,  0.06300388, -0.00223571, -0.15878877,
		%         -0.19725296, -0.24326466])
		\begin{tabular}{l|c|cc}
			\toprule
			\textbf{system}& Data & WERR & SA-WERR   \\ 	
			\midrule
			Original  & N/A & 0 & 0 \\
			BLSTM & 219hr & -6.4\% & -18.8\% \\
			Conformer-base & 219hr & -7.2\% & -6.3\% \\
			Conformer-large& 219hr & -2.5 \% & 1.9 \% \\
			Conformer-base &1500hr & 9.5\% & 8.8\% \\
			Conformer-base-merge& 1500hr & 8.4\% & 10.13\% \\
			Conformer-base-merge-nlabel& 1500hr & 11.8\% & 13.7\% \\
			Conformer-large-merge-nlabel & 1500hr & 8.08\% & 18.4\% \\
			\bottomrule
		\end{tabular}
		
	\end{table}

	Table  \ref{tab:prod} shows the WER and SA-WER reduction rates. 
	With the three improvements described above, the proposed model reduced the WER and SA-WER by  $11.8\%$ and $18.4\%$ relative, respectively, compared with a system without the separation front-end. Although the BLSTM based network improved the recognition result for the LibriCSS dataset especially for the high overlap ratio settings, it largely degraded the speech recognition and speaker diarization performance on the Real Conversation dataset. Because the speech overlap happens only sporadially in real conversations, it is important for the separation model not to hurt the performance for less overlap cases. 
	%We believe two reasons contribute to this degradation. Firstly, as the overlap ratio in real data is less than LibriCSS, the distortion introduced by separation module surpasses the separation benefit it brings. Meanwhile, the real conversation is essentially more complex, creating additional challenges for separation model. 
	Thanks to the better modeling capacity, the Conformer based models significantly mitigates the performance degradation. In addition, it can be seen that each introduced step brought about consistent improvement for both performance metrics. 
	%This results on real data also suggests that a tighter integration across modality potentially leads to further improvement in conversation transcription.  

	% Conformer outperforms BLSTM by 3$\%$ and 8$\%$ in terms of non-overlap and overlap utterance-wise evaluation, respectively, where the evaluation is the WER of a hybrid SR model. 

	\section{Conclusion}
	In this work, we investigated the use of Conformer for continuous speech separation. The experimental results showed that it  outperformed  RNN-based models for both utterance-wise evaluation and continuous input evaluation. 
	The superiority of Conformer to Transformer was also observed. 
	This work is also the first to report 
	substantial WER and SA-WER gains from the speech separation in a single-channel real meeting transcription task. 
	The results indicate the usefulness of appropriately utilizing context information in the speech separation.

	%In the future, we will study how to model long history better. 
	%Besides, how to joint optimize SR system and separation system is another promising direction. 
	%Table~\ref{tab:1ch_cont_result} shows the WER scores for the continuous input evaluation under the single-channel setting. 

	% References should be produced using the bibtex program from suitable
	% BiBTeX files (here: strings, refs, manuals). The IEEEbib.bst bibliography
	% style file from IEEE produces unsorted bibliography list.
	% -------------------------------------------------------------------------
	\bibliographystyle{IEEEbib}
	\bibliography{strings,refs}

\begin{thebibliography}{10}

\bibitem{li2018developing}
Jinyu Li, Rui Zhao, Zhuo Chen, Changliang Liu, Xiong Xiao, Guoli Ye, and Yifan
  Gong,
\newblock ``Developing far-field speaker system via teacher-student learning,''
\newblock in {\em Proc. ICASSP}. IEEE, 2018, pp. 5699--5703.

\bibitem{movsner2019improving}
Ladislav Mo{\v{s}}ner, Minhua Wu, et~al.,
\newblock ``Improving noise robustness of automatic speech recognition via
  parallel data and teacher-student learning,''
\newblock in {\em Proc. ICASSP}. IEEE, 2019, pp. 6475--6479.

\bibitem{sun2019speaker}
Lei Sun, Jun Du, et~al.,
\newblock ``A speaker-dependent approach to separation of far-field
  multi-talker microphone array speech for front-end processing in the chime-5
  challenge,''
\newblock {\em IEEE JSTSP}, vol. 13, no. 4, pp. 827--840, 2019.

\bibitem{Li2020Comparison}
Jinyu Li, Yu~Wu, Yashesh Gaur, Chengyi Wang, Rui Zhao, and Shujie Liu,
\newblock ``On the comparison of popular end-to-end models for large scale
  speech recognition,''
\newblock in {\em Interspeech}, 2020.

\bibitem{watanabe2020chime}
Shinji Watanabe, Michael Mandel, Jon Barker, and Emmanuel Vincent,
\newblock ``Chime-6 challenge: Tackling multispeaker speech recognition for
  unsegmented recordings,''
\newblock {\em arXiv preprint arXiv:2004.09249}, 2020.

\bibitem{yoshioka2019advances}
Takuya Yoshioka, Igor Abramovski, et~al.,
\newblock ``Advances in online audio-visual meeting transcription,''
\newblock in {\em Proc. ASRU}. IEEE, 2019, pp. 276--283.

\bibitem{hershey2016deep}
John~R Hershey, Zhuo Chen, Jonathan Le~Roux, and Shinji Watanabe,
\newblock ``Deep clustering: Discriminative embeddings for segmentation and
  separation,''
\newblock in {\em Proc. ICASSP}. IEEE, 2016, pp. 31--35.

\bibitem{yu2017permutation}
Dong Yu, Morten Kolb{\ae}k, Zheng-Hua Tan, and Jesper Jensen,
\newblock ``Permutation invariant training of deep models for
  speaker-independent multi-talker speech separation,''
\newblock in {\em Proc. ICASSP}. IEEE, 2017, pp. 241--245.

\bibitem{kolbaek2017multitalker}
Morten Kolb{\ae}k, Dong Yu, and Others,
\newblock ``Multitalker speech separation with utterance-level permutation
  invariant training of deep recurrent neural networks,''
\newblock {\em IEEE/ACM TASLP}, vol. 25, no. 10, pp. 1901--1913, 2017.

\bibitem{luo2019conv}
Yi~Luo and Nima Mesgarani,
\newblock ``Conv-tasnet: Surpassing ideal time--frequency magnitude masking for
  speech separation,''
\newblock {\em IEEE/ACM TASLP}, vol. 27, no. 8, pp. 1256--1266, 2019.

\bibitem{luo2020dual}
Yi~Luo, Zhuo Chen, and Takuya Yoshioka,
\newblock ``Dual-path rnn: efficient long sequence modeling for time-domain
  single-channel speech separation,''
\newblock in {\em Proc. ICASSP}. IEEE, 2020, pp. 46--50.

\bibitem{chen2020dual}
Jingjing Chen, Qirong Mao, and Dong Liu,
\newblock ``Dual-path transformer network: Direct context-aware modeling for
  end-to-end monaural speech separation,''
\newblock {\em arXiv preprint arXiv:2007.13975}, 2020.

\bibitem{vaswani2017attention}
Ashish Vaswani, Noam Shazeer, Niki Parmar, Jakob Uszkoreit, Llion Jones,
  Aidan~N Gomez, {\L}ukasz Kaiser, and Illia Polosukhin,
\newblock ``Attention is all you need,''
\newblock in {\em NIPS}, 2017, pp. 5998--6008.

\bibitem{dong2018speech}
Linhao Dong, Shuang Xu, and Bo~Xu,
\newblock ``Speech-transformer: a no-recurrence sequence-to-sequence model for
  speech recognition,''
\newblock in {\em Proc. ICASSP}. IEEE, 2018, pp. 5884--5888.

\bibitem{chang2020end}
Xuankai Chang, Wangyou Zhang, Yanmin Qian, Jonathan Le~Roux, and Shinji
  Watanabe,
\newblock ``End-to-end multi-speaker speech recognition with transformer,''
\newblock in {\em ICASSP}. IEEE, 2020, pp. 6134--6138.

\bibitem{chen2020continuous}
Zhuo Chen, Takuya Yoshioka, et~al.,
\newblock ``Continuous speech separation: Dataset and analysis,''
\newblock in {\em Proc. ICASSP}. IEEE, 2020, pp. 7284--7288.

\bibitem{he2019streaming}
Yanzhang He, Tara~N Sainath, et~al.,
\newblock ``Streaming end-to-end speech recognition for mobile devices,''
\newblock in {\em Proc. ICASSP}. IEEE, 2019, pp. 6381--6385.

\bibitem{zhang2020transformer}
Qian Zhang, Han Lu, Hasim Sak, et~al.,
\newblock ``Transformer transducer: A streamable speech recognition model with
  transformer encoders and {RNN-T} loss,''
\newblock in {\em Proc. ICASSP}, 2020.

\bibitem{gulati2020conformer}
Anmol Gulati, James Qin, et~al.,
\newblock ``Conformer: Convolution-augmented transformer for speech
  recognition,''
\newblock {\em arXiv preprint arXiv:2005.08100}, 2020.

\bibitem{yoshioka2018multi}
Takuya Yoshioka, Hakan Erdogan, Zhuo Chen, and Fil Alleva,
\newblock ``Multi-microphone neural speech separation for far-field
  multi-talker speech recognition,''
\newblock in {\em ICASSP}. IEEE, 2018, pp. 5739--5743.

\bibitem{wang2014training}
Yuxuan Wang, Arun Narayanan, and DeLiang Wang,
\newblock ``On training targets for supervised speech separation,''
\newblock {\em IEEE/ACM TASLP}, vol. 22, no. 12, pp. 1849--1858, 2014.

\bibitem{erdogan2017deep}
Hakan Erdogan, John~R Hershey, Shinji Watanabe, and Jonathan Le~Roux,
\newblock ``Deep recurrent networks for separation and recognition of
  single-channel speech in nonstationary background audio,''
\newblock in {\em New Era for Robust Speech Recognition}, pp. 165--186.
  Springer, 2017.

\bibitem{6516079}
M.~{Souden}, S.~{Araki}, K.~{Kinoshita}, T.~{Nakatani}, and H.~{Sawada},
\newblock ``A multichannel mmse-based framework for speech source separation
  and noise reduction,''
\newblock {\em IEEE/ACM TASLP}, vol. 21, no. 9, pp. 1913--1928, 2013.

\bibitem{shaw2018self}
Peter Shaw, Jakob Uszkoreit, and Ashish Vaswani,
\newblock ``Self-attention with relative position representations,''
\newblock in {\em NAACL}, 2018, pp. 464--468.

\bibitem{ioffe2015batch}
Sergey Ioffe and Christian Szegedy,
\newblock ``Batch normalization: Accelerating deep network training by reducing
  internal covariate shift,''
\newblock in {\em ICML}, 2015, pp. 448--456.

\bibitem{Yoshioka2018Unmix}
Takuya Yoshioka, Hakan Erdogan, Zhuo Chen, Xiong Xiao, and Fil Alleva,
\newblock ``Recognizing overlapped speech in meetings: A multichannel
  separation approach using neural networks,''
\newblock in {\em Interspeech}, 2018, pp. 3038--3042.

\bibitem{dai2019transformer}
Zihang Dai, Zhilin Yang, and Others,
\newblock ``Transformer-xl: Attentive language models beyond a fixed-length
  context,''
\newblock {\em arXiv preprint arXiv:1901.02860}, 2019.

\bibitem{wang2019semantic}
Chengyi Wang, Yu~Wu, Yujiao Du, Jinyu Li, Shujie Liu, Liang Lu, Shuo Ren, Guoli
  Ye, Sheng Zhao, and Ming Zhou,
\newblock ``Semantic mask for transformer based end-to-end speech
  recognition,''
\newblock {\em Interspeech}, 2020.

\bibitem{wsj1}
Linguistic Data~Consortium Philadelphia,
\newblock ``{CSR-II (WSJ1) Complete},'' 1994,
\newblock \url{http://catalog.ldc.upenn.edu/LDC94S13A}.

\bibitem{imagemethod}
J.~Allen and D.~Berkley,
\newblock ``Image method for efficiently simulating small-room acoustics,''
\newblock {\em JASA}, vol. 65, pp. 943--950, 1979.

\bibitem{habets2007generating}
Emanu{\"e}l~AP Habets and Sharon Gannot,
\newblock ``Generating sensor signals in isotropic noise fields,''
\newblock {\em JASA}, vol. 122, no. 6, pp. 3464--3470, 2007.

\bibitem{loshchilov2018decoupled}
Ilya Loshchilov and Frank Hutter,
\newblock ``Decoupled weight decay regularization,''
\newblock in {\em ICLR}, 2018.

\end{thebibliography}
	
\end{document}